\documentclass[structabstract]{aa}  
\usepackage{graphicx}
\usepackage{txfonts}
\begin{document}
   \title{Physical properties of galaxies and their evolution\\ in the VIMOS VLT Deep Survey\thanks{based on data obtained
with the European Southern Observatory Very Large Telescope, Paranal, Chile, program 070.A-9007, and on data obtained
at the Canada-France-Hawaii Telescope, operated by the CNRS in France, CNRC in Canada and the University of Hawaii.}}
\subtitle{II. Extending the mass-metallicity relation to the range $z \approx 0.89 -1.24$}
\titlerunning{Extending the M-Z relation in VVDS up to $z \approx 1.24$}

  \author{E. P\'erez-Montero\inst{1}\thanks{Postdoctoral fellow of the Spanish {\em Fundaci\'on Espa\~nola para la Ciencia
y la Tecnolog\'\i a}.}
 \and T. Contini \inst{1}
 \and F. Lamareille \inst{1}
 \and J. Brinchmann \inst{19}
\and C. J. Walcher \inst{7}
\and S. Charlot \inst{8,10}
\and M. Bolzonella  \inst{3} 
\and L. Pozzetti    \inst{3} 
\and D. Bottini \inst{2}
\and B. Garilli \inst{2}
\and V. Le Brun \inst{7}
\and O. Le F\`evre \inst{7}
\and D. Maccagni \inst{2}
\and R. Scaramella \inst{4,13}
\and M. Scodeggio \inst{2}
\and L. Tresse \inst{7}
\and G. Vettolani \inst{4}
\and A. Zanichelli \inst{4}
\and C. Adami \inst{7}
\and S. Arnouts \inst{23,7}
\and S. Bardelli  \inst{3}
\and A. Cappi    \inst{3}
\and P. Ciliegi    \inst{3}  
\and S. Foucaud \inst{21}
\and P. Franzetti \inst{2}
\and I. Gavignaud \inst{12}
\and L. Guzzo \inst{9}
\and O. Ilbert \inst{20}
\and A. Iovino \inst{9}
\and H.J. McCracken \inst{10,11}
\and B. Marano     \inst{6}  
\and C. Marinoni \inst{18}
\and A. Mazure \inst{7}
\and B. Meneux \inst{2,9}
\and R. Merighi   \inst{3} 
\and S. Paltani \inst{15,16}
\and R. Pell\`o \inst{1}
\and A. Pollo \inst{7,17}
\and M. Radovich \inst{5}
\and D. Vergani \inst{2}
\and G. Zamorani \inst{3} 
\and E. Zucca  \inst{3}        }

 \offprints{E. P\'erez-Montero}

  \institute{\center{ ({\em Affiliations can be found after the references} )}}

   \date{}  

 \abstract
{}
{We present a continuation of our study about the relation between stellar mass and gas-phase metallicity in the VIMOS VLT Deep Survey (VVDS). In this work we extend the determination of metallicities up to redshift $\approx$ 1.24 for a sample of 42 star-forming galaxies with a mean redshift value of 0.99.}
{For a selected sample of emission-line galaxies, we use both diagnostic diagrams and empirical calibrations based on [O{\sc ii}] emission lines along with the empirical relation between the intensities of the [O{\sc iii}] and [Ne{\sc iii}] emission lines and the 
theoretical ratios between Balmer recombination emission lines to identify star-forming galaxies and to derive their metallicities. 
We derive stellar masses by fitting the whole spectral energy distribution with a set of stellar population synthesis models.}
{These new methods allow us to extend the mass-metallicity relation to higher redshift. We show that
the metallicity determinations are consistent with more established strong-line methods. Taken together
this allows us to study the evolution of the mass-metallicity relation
up to z $\approx$ 1.24 with good control of systematic uncertainties. We find an evolution with redshift of the average metallicity 
of galaxies very similar to those reported in the literature: for a given stellar mass, galaxies at $z \sim 1$ have, on average, 
a metallicity $\sim 0.3$ dex lower than galaxies in the local universe. However we do not see any significant metallicity evolution 
between redshifts $z \sim 0.7$ (Paper I) and $z \sim 1.0$ (this paper). We find also the same flattening of the mass-metallicity 
relation for the most massive galaxies as reported in Paper I at lower redshifts, but again no apparent evolution of the slope 
is seen between $z \sim 0.7$ and $z \sim 1.0$.}
{} 

        
  \maketitle
 
%

\section{Introduction}
The formation and evolution of galaxies at different cosmological epochs are driven mainly by two processes: the star formation history and the metal enrichment. These processes, which are strongly connected in the evolution of individual galaxies, can be altered by the peculiar dynamics of the gas. The rate of star formation can be increased as a consequence of mergers or due to the inflow of gas. Regarding metallicity, it is sensitive to metal losses due to stellar winds, supernovae and active nucleus feedbacks.  Therefore, the study of the dominant mechanisms of evolution of galaxies at different redshifts rely on the determination of metallicity and total stellar mass.

The correlation between stellar mass and metallicity (M-Z relation) was firstly found in the nearby universe by Lequeux et al. (1979) for dwarf irregular galaxies. Since them, numerous works have been performed in order to understand and quantify this relation in the local universe (among others Skillman et al., 1989; Brodie \& Huchra, 1991; Zaritsky et al.,
1994; Richer \& McCall, 1994; Garnett et al., 1997 or Pilyugin \& Ferrini, 2000). This relation has its counterpart in
the luminosity-metallicity relation which is easier to derive when a lower number of spectrophotometric bands are
available. Besides, in the last years, this work
has been complemented with the 
 use of large surveys of galaxies which allow the derivation of this relation in samples
statistically more significant, ({\em e.g.} SDSS, Tremonti et al., 2004; 2dFGRS, Lamareille et al., 2004), that confirm the decrease of metal loss when galaxy stellar mass increases. Several studies of the evolution of this relation at larger redshifts have appeared due to the growing number of spectroscopic surveys of star-forming galaxies in the distant universe. This is the case of Savaglio et al. (2005) or Lamareille et al. (2006) who agree to find lower mean metallicities for the same galaxy mass bins at larger redshifts. However, they do not agree about the evolution of the slope of the M-Z relation: Lamareille et al. (2006) reported a flattening of the relation at higher redshifts whereas Savaglio et al. (2005) found a steepening of the M-Z relation.
Unfortunately, these studies are limited to $z < 1$ galaxies, due to the required presence of certain emission lines in the optical spectrum to derive the metallicity.  Erb et al. (2006) studied this relation for a sample of star forming galaxies at $z \approx 2$ using near-infrared spectra and they find a lower value of the mean metallicity for a given galaxy stellar mass.
This evolution is apparently much more stronger at higher redshift as reported by Maiolino et al. (2008), using also IR spectra for a sample of star-forming galaxies ar z$>$3.
 However, there is a lack of observations and available data in the redshift range $0.9 - 2$, corresponding to the epoch of the universe when the maximum of star formation took place. It is therefore particularly interesting to study the M-Z relation at this period.

The estimation of the global trend of the M-Z relation at different cosmological epochs is possible only through the analysis
of statistically significant samples with accurate determinations of the stellar mass and the gas phase oxygen abundance.
This paper is a continuation of Lamareille et al. (2008; hereinafter Paper I), who study the M-Z relation of the star-forming galaxies detected in the VIMOS VLT Deep Survey up to redshift $0.89$. In this work we extend the upper redshift limit of the study up to $z \approx 1.24$, using the relation between metallicity and emission-line strength of [O{\sc ii}] and [Ne{\sc iii}] presented in P\'erez-Montero et al. (2007). In the next section we describe the sample of star-forming galaxies discussing the selection criteria and the methods to derive their stellar mass and metal content, which can go
to higher redshift than other classical methods. In section 3, we describe the M-Z relation in the redshift bin $0.89 < z <1.24$, with an average value in our sample of $z$ = 0.99 and we compare our results with those found in Paper I. Finally, we summarise our results and present our conclusions in the last section. Throughout this paper we normalize
the derived stellar masses and the absolute magnitudes with the standard $\Lambda$-CDM cosmology, {\em i.e.},
$h$ = 0.7, $\Omega_m$ = 0.3 and $\Omega_\Lambda$ = 0.7 (Spergel et al., 2007).

\section{Description of the sample}
The VIMOS/VLT Deep Survey (VVDS, Le F\`evre et al., 2005) is one of the widest and deepest spectrophotometric surveys of distant galaxies with a mean redshift of $z \approx 0.7$. The optical spectroscopic data belong to the first epoch sample obtained with the VIsible MultiObject Spectrograph (VIMOS) at ESO/VLT (UT3). They are divided in two deep fields ($17.5 \leq$ I$_{AB}$ $\leq 24$): VVDS-02h (hereafter F02; Le F\`evre et al., 2005) and CDFS (Le F\`evre et al., 2004) and in three wide fields ($17.5 \leq$ I$_{AB}$ $\leq 22.5$): VVDS-10h, VVDS-14h and VVDS-22h (hereafter F10, F14 and F22 respectively, Garilli et al., in prep.). 
As explained in Paper I, we divide therefore our
total sample in two: a deep one with the F02 and CDFS fields and a wide one with F10, F14 and F22 fields.
Spectroscopic data have been reduced using the {\em VIPGI} pipeline (Scodeggio et al., 2005).

The following selection criteria have been applied on this sample. We consider only objects with a redshift known at a 75\% confidence 
level ({\em i.e.} VVDS redshift flags 2, 3, 4 and 9) and we remove all stars, broad-line Active Galactic Nuclei (hereafter AGNs; 
Gavignaud et al., 2006) and duplicate observations.

\begin{figure}
   \centering
   \includegraphics[width=8cm,clip=]{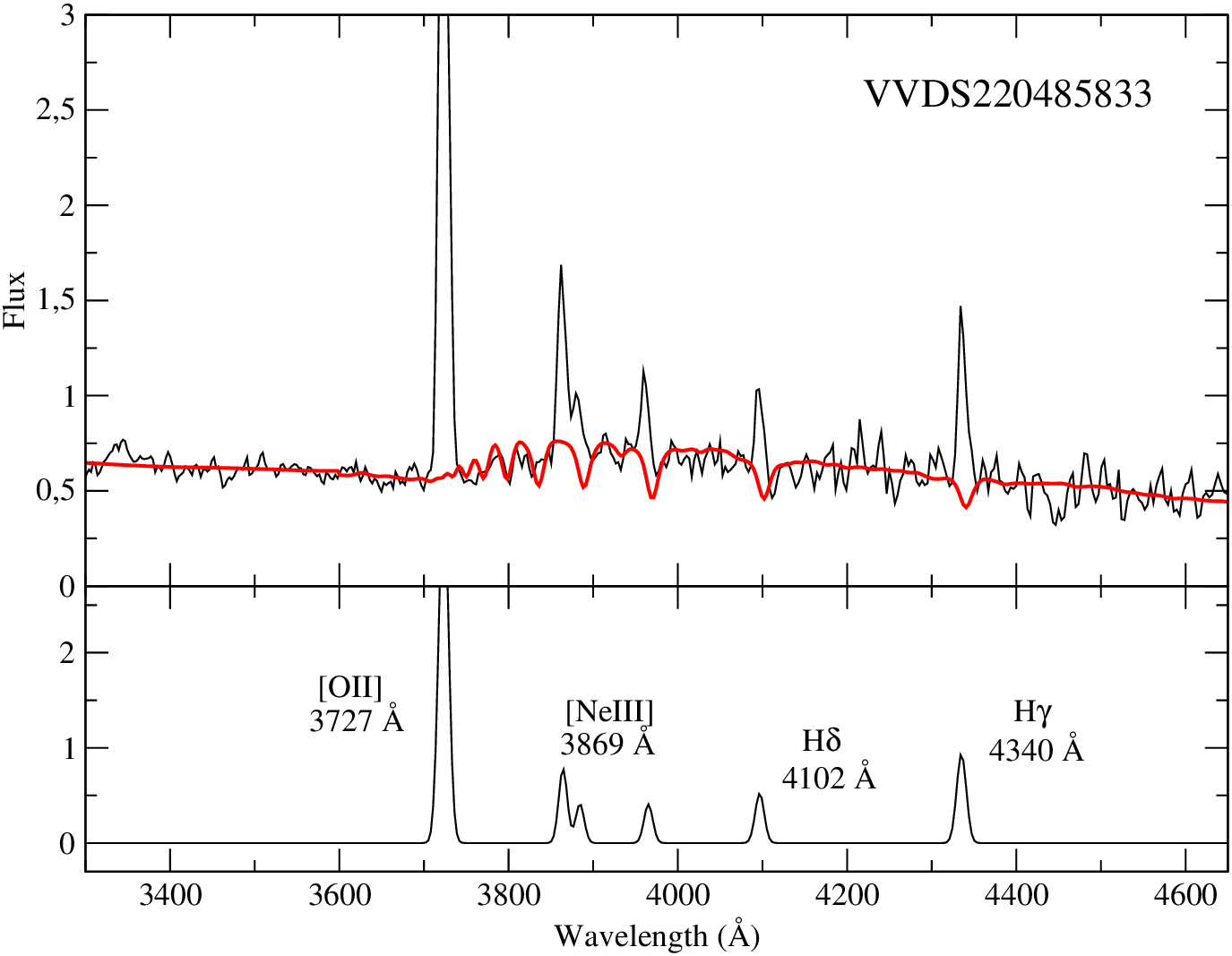}

   \caption{VIMOS rest-frame spectrum of the galaxy VVDS220485833. The top panel illustrates the fitting of the stellar continuum obtained with the {\em platefit\_vimos} tool. In the bottom panel, we show the resultant nebular emission-lines spectrum, once the fitted stellar continuum has been substracted. The labels indicate the set of emission lines used in this work to derive metallicities.}
              \label{spectrum}
    \end{figure}

The optical spectra of each galaxy have been split into two components (stellar and nebular) using the pipeline 
{\em platefit\_vimos} as described in Paper I. This software removes the stellar continuum and absorption lines and then fits automatically all
emission lines. The templates used to fit the stellar component have been derived from the libraries of Bruzual \& Charlot (2003)
resampled to the velocity dispersion of VVDS spectra. The fitting and measurement procedures of all emission lines (flux and equivalent width) 
and stellar indices in the original spectra are described in Paper I. 
In Figure \ref{spectrum} we show an example of the fitting of the stellar continuum (top panel) made by {\em platefit\_vimos} for the galaxy VVDS221485833, {\bf one of the best S/N spectra}. The resultant nebular emission-lines spectrum, once this fitting is substracted, is shown in the bottom panel.

 
   \begin{figure}[t]
   \centering
   \includegraphics[width=8cm,clip=]{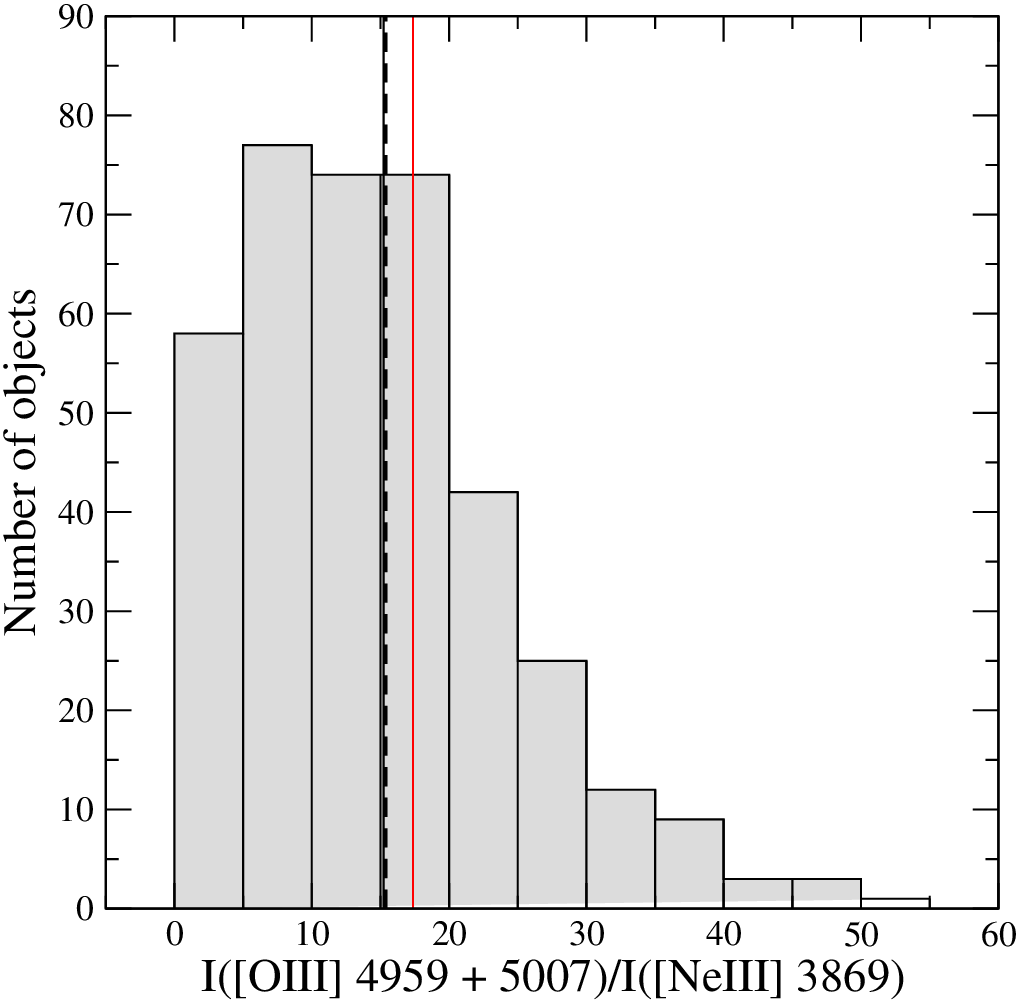}
   \caption{Histogram of the distribution of the ratio between the deredenned fluxes of [O{\sc iii}]4959$+$5007{\AA} and 
[Ne{\sc iii}]3869{\AA} emission lines for the five VVDS fields (VVDS-02h, VVDS-10h, VVDS-14h, VVDS-22h and CDFS). 
The mean value of this distribution, shown as a black solid line, is 15.23. The red solid line represents the mean value of the 
distribution without applying the mean reddening correction and the dashed line is mean value ($= 15.37$) of this emission-line ratio for a 
representative sample of nearby galaxies studied in P\'erez-Montero et al. (2007).}
              \label{o3ne3}
    \end{figure}

 
\subsection{Selection of star-forming galaxies}

In order to  correctly study the M-Z relation, we select the galaxies whose main mechanism of ionization is star formation. 
We use emission-line diagnostic diagrams to identify narrow-line AGNs (Seyfert 2 and LINERS) once broad line AGNs have been
removed. In particular we use the ``blue'' diagnostic diagram based on [O{\sc iii}]5007{\AA}, 
H$\beta$ and [O{\sc ii}]3727{\AA} emission lines (Lamareille et al., 2004) and applied in Paper I for the redshift bin $0.5 < z < 0.9$.
We extend this analysis up to redshift $\sim 1.3$ using the theoretical ratios between H$\beta$ and the most intense 
Balmer hydrogen recombination lines at bluer wavelengths. This is the case of H$\gamma$ at 4340{\AA} and H$\delta$ at 4102{\AA}.
We assume the following values for these ratios, typical of star-forming galaxies:
 H$\gamma$/H$\beta$ = 0.48 and H$\delta$/H$\beta$ = 0.25.

\begin{table*}[t]
\caption{Distribution of star-forming galaxies and narrow-line AGNs among VVDS emission-line galaxies using the ``blue'' diagnostic diagrams
based on the relation between [O{\sc iii}] (blue, $0.5 < z < 0.9$) or [Ne{\sc iii}] (blue-Ne, $0.5 < z < 1.24$) and [O{\sc ii}] emission lines. 
We show the distribution for the whole sample and for wide and deep fields. In the wide classification we include those galaxies belonging
to the deep fields having $I_{AB} \leq$ 22.5 }
\centering
\begin{tabular}{l c c c c c c}
\hline
\hline
 &  \multicolumn{2}{c}{0.5 $<z<$ 0.9} & \multicolumn{2}{c}{0.5 $<z<$ 0.9} & \multicolumn{2}{c}{$z>$ 0.9} \\
Sample/Diagnostic & blue & \% & blue-Ne & \% & blue-Ne & \% \\
\hline
{\bf Wide}         \\
Emission-line (total) &  924 & {\em 100} & 338 & {\em 100} & 62 & {\em 100} \\
Star-forming    &  768 & {\em 83}  & 157  & {\em 45} & 34 & {\em 55}\\  
Candidates    &  138 &  {\em 15} & 142  & {\em 42} & 27 & {\em 43}\\
Seyfert 2      &    18 &  {\em 2} & 39  & {\em 12} & 1 & {\em 2}\\
LINER      &      0  &   {\em 0} &  0  &   {\em 0} &  0  &   {\em 0} \\
\hline
{\bf Deep} \\
Emission-line (total) &  568 & {\em 100} & 197 & {\em 100} & 36 & {\em 100} \\
Star-forming    &  412 & {\em 73} & 110  & {\em 56} & 18 & {\em 50}  \\
Candidates      &  139 & {\em 24} & 79  & {\em 40} & 17 & {\em 47} \\ 
Seyfert 2      &  17   &  {\em 3} & 8  & {\em 4} & 1 & {\em 3}\\  
LINER       &      0  &   {\em 0} &  0  &   {\em 0} &  0  &   {\em 0} \\
\hline
{\bf Total} \\
Emission-line (total) &  1213 & {\em 100} & 418 & {\em 100} & 72 & {\em 100} \\
Star-forming    &  945 & {\em 78} & 216  & {\em 52} & 40 & {\em 55}\\  
Candidates    &  236 & {\em 20} & 163  & {\em 39} & 30 & {\em 42}\\  
Seyfert 2      &  32 & {\em 3} & 39  & {\em 9} & 2 & {\em 3}\\  
LINER      &      0  &   {\em 0} &  0  &   {\em 0} &  0  &   {\em 0} \\
\hline
\hline

\end{tabular}
\label{agns}
\end{table*} 


\begin{figure*}
   \centering
   \includegraphics[width=14cm,clip=]{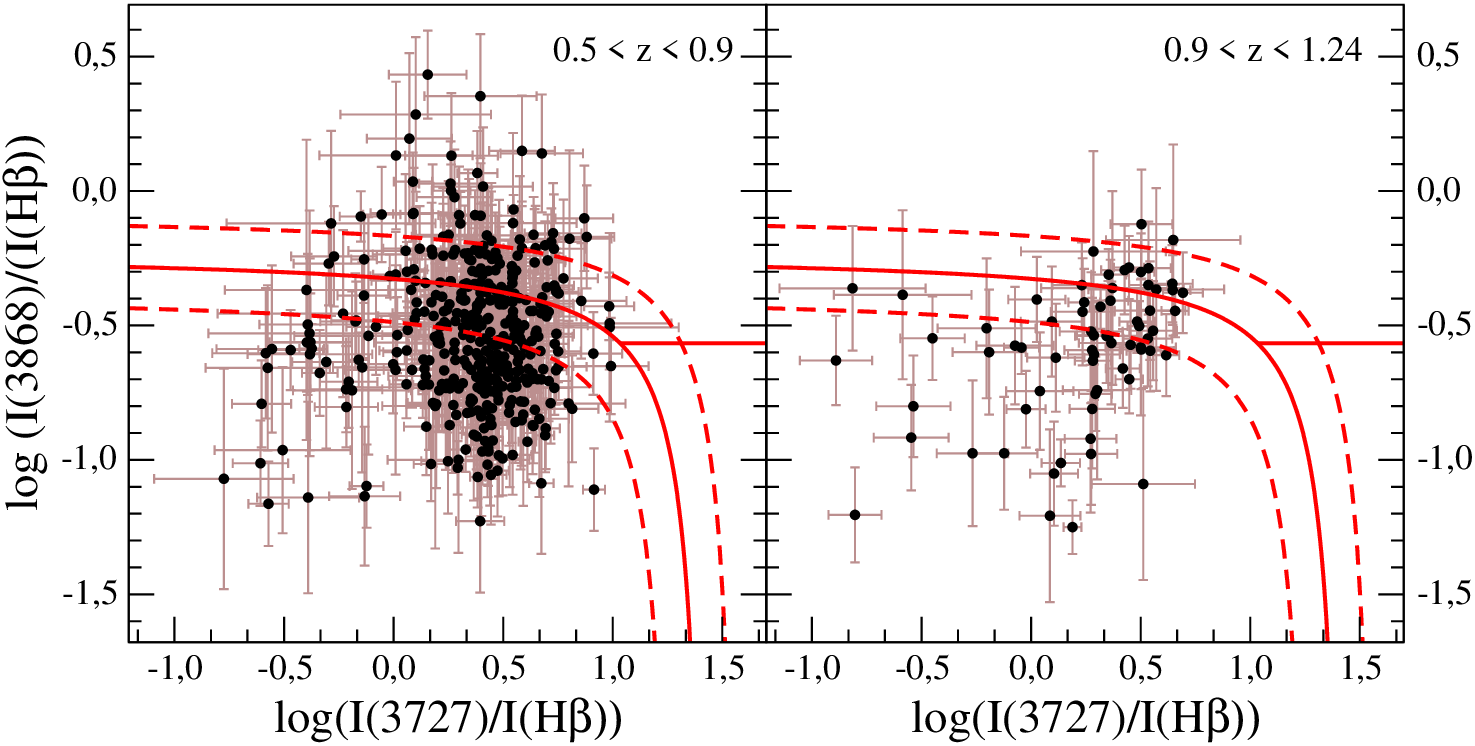}
      \caption{Spectral classification of VVDS galaxies using the diagram diagnostic based
on [Ne{\sc iii}] emission line. In the left panel we show the diagram for 418 emission-line galaxies in the redshift range $0.5 < z < 0.9$, while the right panel shows 72 emission-line galaxies in the redshift range $0.9 < z < 1.24$. In these two panels, the red solid curve divides the regions of star-forming galaxies and narrow-line AGNs. The straight red solid line divides the AGN region between Seyfert 2 (top) and LINER (bottom).}
              \label{diagblue}
    \end{figure*}


Although these ratios slightly depend on the electron temperature of the ionized gas, this dependence is negligible in comparison with the intrinsic uncertainty of the measurement of the involved emission lines, even in the case
of narrow-line AGNs, which show the lowest values of these ratios. 
We apply the empirical relation found between the emission-line ratio of [O{\sc iii}]4959$+$5007{\AA} and [Ne{\sc iii}]3869{\AA} (P\'erez-Montero et al., 2007). The mean value of this ratio reported in this study for a sample of HII regions of the local universe is equal to 15.37. However, the mean value of the same ratio for the VVDS star-forming galaxies (both for the deep and wide fields) is equal to 17.37. We correct this slight difference taking into account a reddening correction corresponding to an extinction of A$_V = 0.5$ mag. After this correction, the mean value of [O{\sc iii}]4959$+$5007{\AA}/[Ne{\sc iii}]3869{\AA} emission-line ratio for the VVDS galaxies is 15.23, which is almost identical to the value found for the reddening corrected HII regions of the local universe. 
Therefore, we decide to use the same reddening correction for the rest of the calculations in the VVDS sample.
In Figure \ref{o3ne3}, we show the distribution of the emission-line ratio [O{\sc iii}]4959$+$5007{\AA}/[Ne{\sc iii}]3869{\AA} for the 379 VVDS secure and candidates star-forming galaxies for which we have measured these lines in the redshift range $0.5 < z < 0.9$ and with an error on the ratio lower than the ratio itself. In this figure we compare the mean value of this distribution, once extinction corrected (black solid line), the mean value found without extinction correction (red solid line) and the value found in the local universe by P\'erez-Montero et al. (2007, dashed line).


   \begin{figure*}
   \centering
   \includegraphics[width=14cm,clip=]{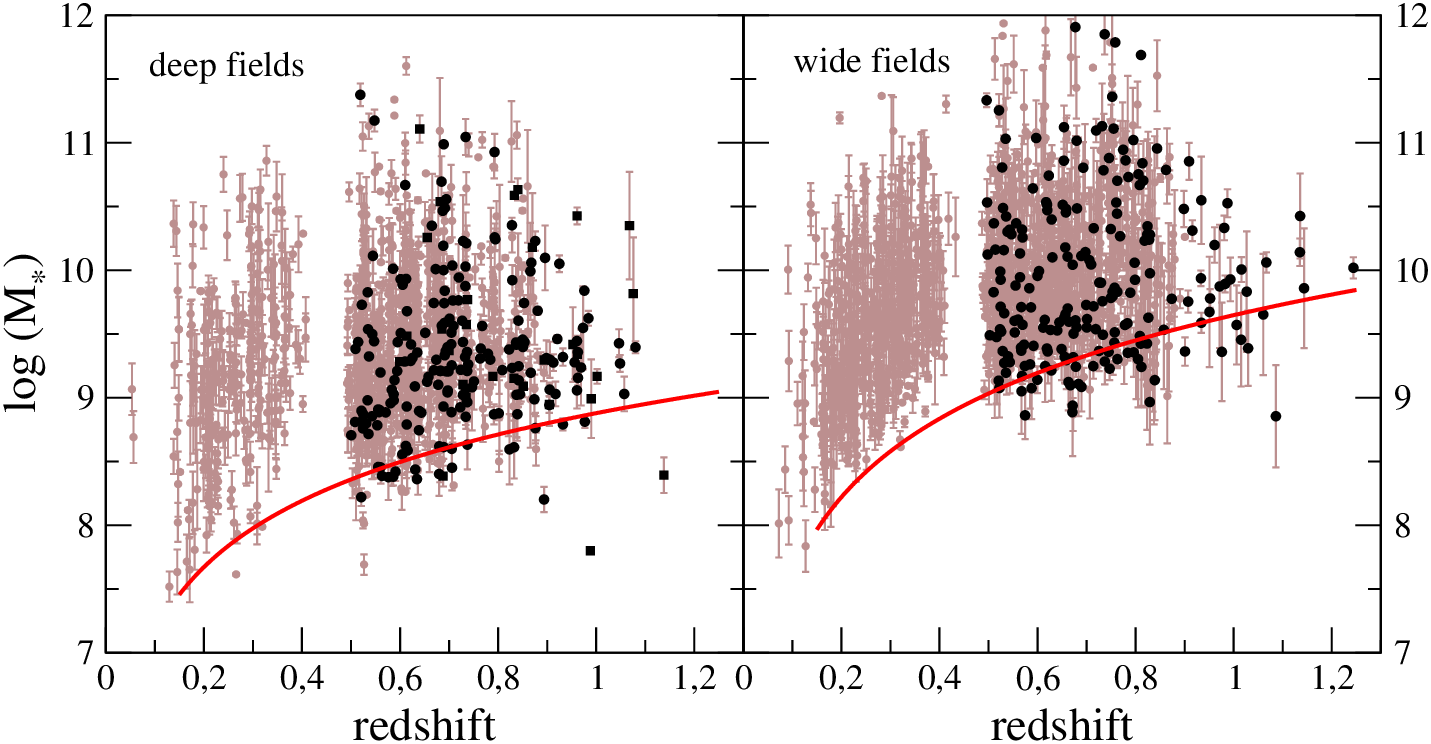}
   \caption{Stellar masses of star-forming galaxies (in solar masses) in the VVDS deep (left panel) and wide (right panel) fields 
as a function of redshift. The squares represent the star-forming galaxies selected using the diagnostic 
diagram based on [Ne{\sc iii}] emission line.
The solid curve represents a logarythmic fit to the 50\% mass-to-light completude levels and
shows the bias imposed by the VVDS selection on apparent magnitude.}
              \label{zM}
    \end{figure*}


The empirical limit between star-forming galaxies and narrow-line AGNs in the ``blue-Ne'' diagnostic diagram based on [Ne{\sc iii}]3869{\AA} 
emission line takes the following form:
\begin{equation}
\log \left(\frac{[NeIII]}{H\beta}\right) = \frac{0.14}{log ([OII]/H\beta) - 1.45} - 0.23
\end{equation}

Following the same prescriptions as in the ``blue'' diagnostic method, the objects lying under this line are 
considered as star-forming galaxies, while those lying above it are considered as AGNs.  This 
diagnostic diagram is, in fact, equivalent to that proposed by Rola et al. (1997) which
is based on the same emission lines.
In the AGN zone, LINERs are identified by the following criterium:
\begin{equation}
\log\left(\frac{I([NeIII] 3869}{I(H\beta)}\right) < - 0.57
\end{equation}

Adopting the emission-line ratios H$\gamma$/H$\beta$ and H$\delta$/H$\beta$ listed above, we can use the same expression for galaxies at redshifts higher than 0.9 or for those whose H$\beta$ emission line is not measured in the redshift range $0.5 < z < 0.9$. 
In the case of H$\delta$, these expressions take the same form as the equations (23) and (24) in P\'erez-Montero et al. (2007). 

In Figure \ref{diagblue} we show the [O{\sc ii}]/H$\beta$ vs. [Ne{\sc iii}]/H$\beta$ relation 
for the sample of VVDS galaxies with the appropriate lines, in the redshift ranges $0.5 < z < 0.9$
(left panel) and $0.9 < z < 1.24$ (right panel).
The dashed lines delineate the uncertainty domain ($\pm 0.15$ dex around the empirical limit) in the spectral classification: objects lying in this region are considered as ``candidates'' to star-forming galaxies.
The results of this analysis are summarised in Table \ref{agns}. We make also a comparison with the classification 
derived from the ``blue'' diagnostic diagram based on [O{\sc ii}] and [O{\sc iii}] emission lines for galaxies 
in the redshift range $0.5 < z < 0.9$.  

We see that the fraction of galaxies classified as candidates to star-forming from the diagram based on 
[Ne{\sc iii}] is much higher (at all redshift bins) that the one based on [O{\sc iii}] emission line. Taking into account 
that the rate of coincidence between blue and blue-Ne diagnostics for the galaxies for which it is possible to apply at the 
same time the two methods reaches 98\%, we think that the sample of galaxies with accurate [Ne{\sc iii}] measurements (i.e. with a high enough S/N) 
is somehow biased towards a population with enhanced [Ne{\sc iii}] emission. 
This is indicated by the range of log([Ne{\sc iii}]/H$\beta$) values 
in both panels of Figure \ref{diagblue} which are between -1 and 0, while in Fig. 3 of Paper I we see that 
the ratio log([O{\sc iii}]/H$\beta$), which is approximately an order of magnitude larger
than log([Ne{\sc iii}]/H$\beta$), varies between -0.5 and 1. 
This means that in our sample we are certainly affected by some selection effect that
preserves us to properly detect all the star forming galaxies with a value of 
log([Ne{\sc iii}]/H$\beta$) between -1.5 and -1. Therefore, the relative number of candidates is larger
in this subsample.

This same effect is seen when looking at the relative number of AGNs in the redshift bin $0.5 < z < 0.9$, especially in the wide 
VVDS sample. On the other hand, when we compare the relative numbers of AGNs in the redshift regime $z > 0.9$, we do not see a 
significant increase compared with the blue diagnostic diagram at lower redshift. Note, finally, that we did not identify any LINER 
in our VVDS sample.

For the following analysis of stellar masses and metallicities, we consider both the samples of candidates and secure star-forming 
galaxies. 
Unfortunately, there are no deep enough X-ray or far-IR observations of the VVDS fields to check the presence of
hidden AGNs.

\subsection{Estimation of stellar masses}

The procedure to derive stellar masses is based on the comparison between observations, basically the 
photometric Spectral Energy Distribution (hereafter SED) of galaxies, combined in some cases with their 
stellar absorption index H$\delta$ and D4000 break measured in VIMOS spectra, and population synthesis models. 
As described in Paper I and introduced by SDSS collaborators (Kauffmann et al., 2003; Tremonti et al., 2004, Brinchmann et al., 2005), 
we are using the Bayesian approch. 
To summarise, the population synthesis models are based on Bruzual \& Charlot (2003), with the addition of secondary bursts to 
the standard declining exponential star formation history (stochastic library, Salim et al., 2005; Gallazzi et al., 2005).
The SED of the models are compared with the available photometry for VVDS galaxies: 
CFH12k observations in $BVRI$ bands for F02, F22 and F10 fields, and in $I$ band for F14 fields
(McCracken et al, 2003; Le F\`evre et al., 2004), completed by $U$ band observations in the F02 field
(Radovich et al., 2004), CFHTLS (Canada-France-Hawaii Telescope Legacy Survey) in $u^*g'r'i'z'$ bands
for F02 and F22 fields and $JKs$ photometry available for a substantial fraction of the objects
in the F02 field (Iovino et al., 2005; Temporin
et al., 2006). For the CDFS field, we use CFH12k $UBVRI$ (Arnouts et al., 2001) and HST $bviz$ (Giavalisco et al., 2004)
observations. Finally, objects in F02 and F22 fields have been cross-matched with the UKIDSS public
catolog (Warren et al., 2007), providing additional observations in $JK$ bands.


\begin{figure}
   \centering
   \includegraphics[width=8.5cm,clip=]{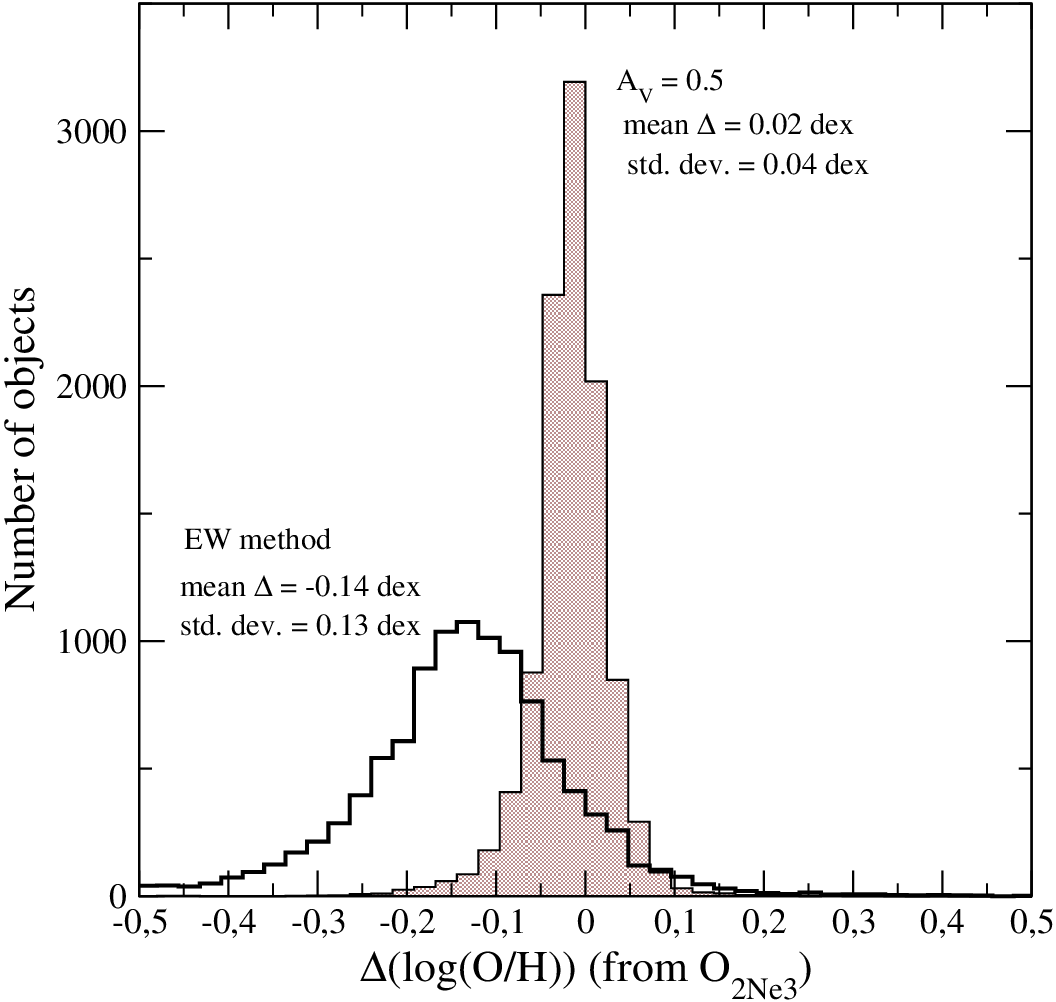}
 
      \caption{Comparison between O/H abundances derived for the star-forming galaxies of the SDSS DR4 catalogue
using the O$_{2Ne3}$ parameter under different assumptions about reddening. Filled histogram represents the difference
between the abundance using emission lines corrected for reddening using the Balmer decrement and assuming an
average reddening corresponding to A$_V$ = 0.5 mag. The thick line histogram represents the same difference but in
relation with the abundances derived from the equivalent widths. We indicate also the mean value and the standard
deviation of the two distributions.}
              \label{ewa}
    \end{figure}



   \begin{figure}
   \centering
   \includegraphics[width=7cm,clip=]{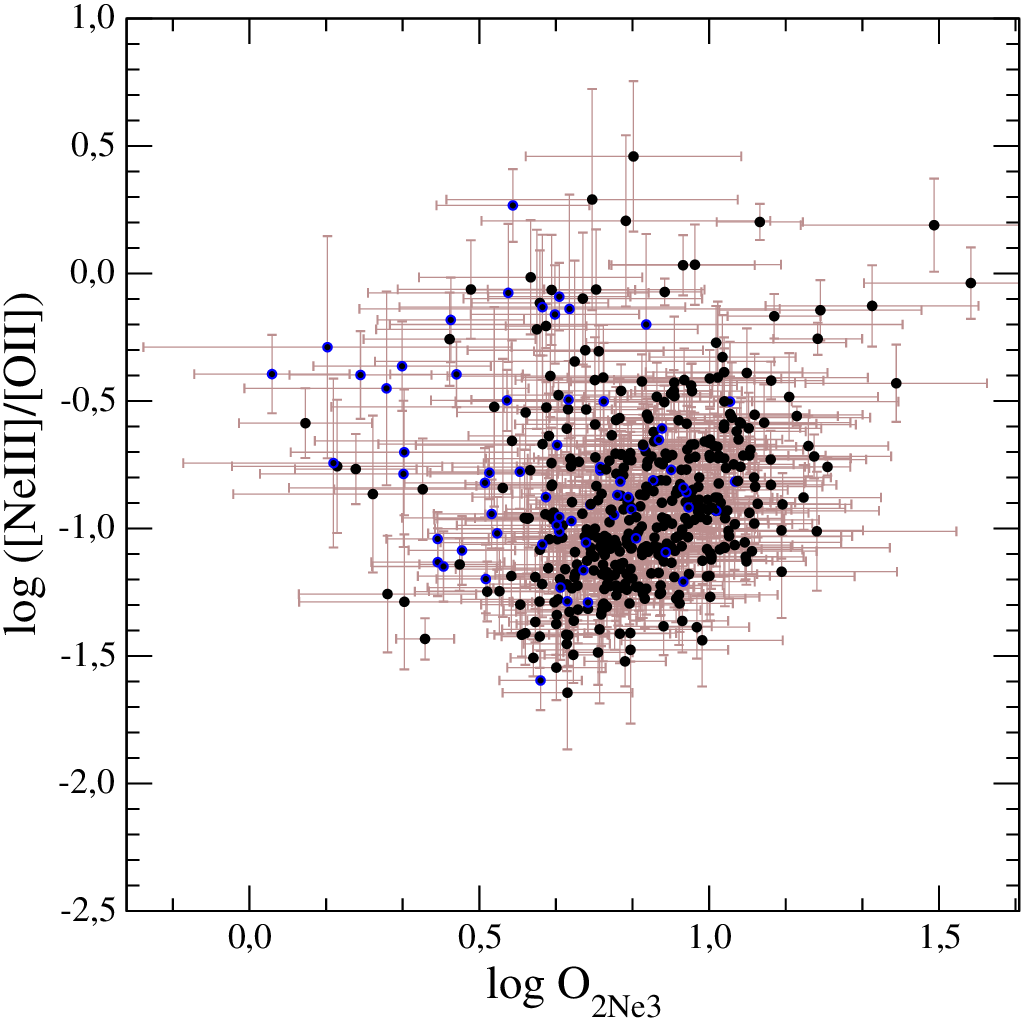}
      \caption{Relation between the O$_{2Ne3}$ parameter and the emission-line ratio [NeIII]/[OII] for the VVDS star-forming galaxies 
at $0.9 < z < 1.24$. O$_{2Ne3}$ has been calculated using either the flux of H$\beta$ (blue squares), H$\gamma$ (black squares), or 
H$\delta$ (white squares).}
              \label{logU}
    \end{figure}


As in Paper I, we correct for the fact that both minimum and maximal stellar detected masses depend
on redshift. We thus define a limiting mass for each redshift bin, which is the minimum mass that would be detected
at a given redshift for a given limiting apparent magnitude. For our M-Z relation study we do not consider galaxies with a 
stellar mass lower that this minimum mass derived in the corresponding redshift bin.

The limiting mass for the redshift range $0.89 <z< 1.24$ is derived empirically using the same procedure as in Paper I.
We derive the limiting mass for each galaxy following the expression:
\begin{equation}
\log(M^*_{lim}) = \log(M^*) + 0.4 \cdot (I_{obs} - I_{sel})
\end{equation}
where $M^*_{lim}$ is the limiting mass in solar masses, $M^*$ the stellar mass in solar masses and $I_{obs}$ is the observed
$I$-band magnitude, $I_{sel}$ is the $I$-band magnitude limit for each field, which is equal to 24 and 22.5 in the case of deep and wide 
fields respectively. In Figure \ref{zM} we show the stellar mass of the selected star-forming galaxies as a function of redshift 
in the range $0 < z < 1.24$. 
The solid curve shows the logarithmic fit to the 50\% mass-to-light ratio completude level of the distribution of the limiting
mass in each redshift bin. Since the average redshift value for the analysed galaxies in the range $0.89 < z < 1.24$ is 
$\overline{z}$ = 1.0, we take a value for the limiting mass in this regime of log(M$_*$/M$_\odot$) = 8.9 in the deep fields 
and 9.6 in the wide fields.
As in Paper I, we obtain a lower value of the limiting mass in the deep fields. The values are also slightly larger than those
derived for the star-forming galaxies in the redshift bin $0.7 < z < 0.9$ which are 8.8 and 9.5 respectively. This is due to
the fact that limiting mass increases with redshift in the selection function.

   \begin{figure*}[t]
   \centering
   \includegraphics[width=14cm,clip=]{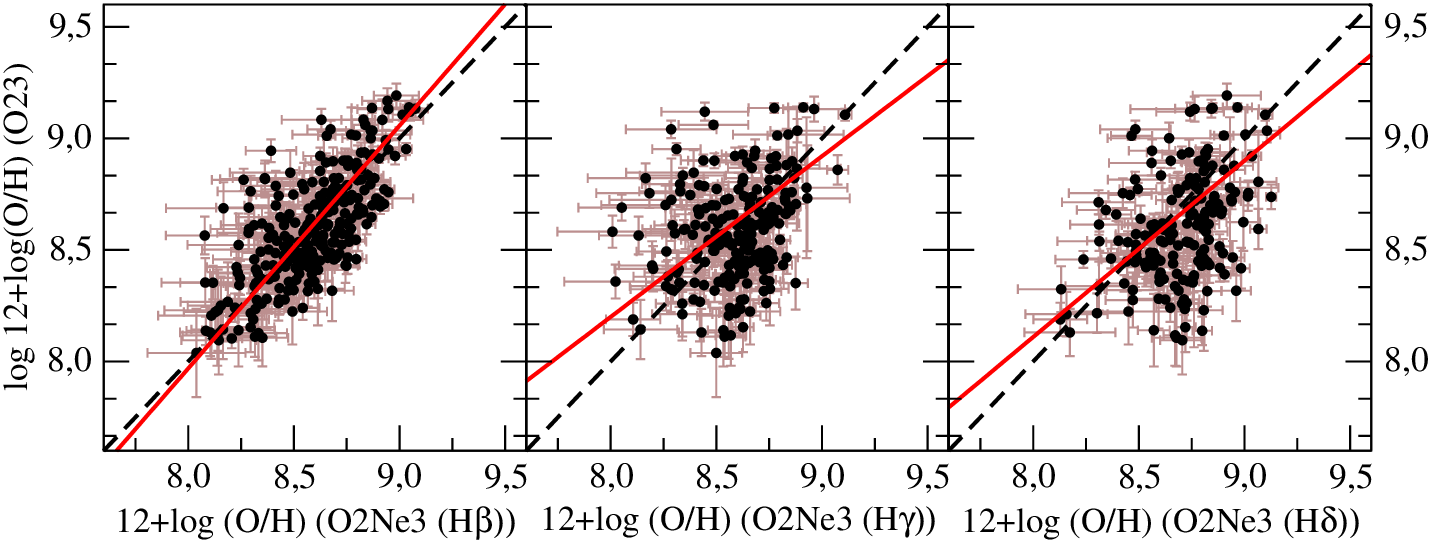}
      \caption{Comparison between the oxygen abundances derived with the O$_{23}$ and O$_{2Ne3}$ parameters. Three cases are distinguished: 
O$_{2Ne3}$ parameter is estimated using H$\beta$ (left panel, 268 galaxies), H$\gamma$ (middle panel, 217 galaxies) or H$\delta$ (right panel, 
187 galaxies) emission line. Only VVDS star-forming galaxies with an accuracy on O/H lower than 0.3 dex are considered. Dashed lines represent 
the 1:1 relation while the solid lines are the best linear fit to the data.}
 \label{ohcomp}
    \end{figure*}

\subsection{Estimation of metallicities}

We estimate metallicities as the oxygen abundances of the ionized gas phase from
the O$_{2Ne3}$ parameter (P\'erez-Montero et al., 2007). This parameter is equivalent
to O$_{23}$ (also known as R$_{23}$), but it is based on the empirical relation existing
between [O{\sc iii}] and [Ne{\sc iii}] emission lines:
\begin{equation}
O_{2Ne3} = \frac{I([OII] 3727 \AA) + 15.37 \cdot I([NeIII] 3869 \AA)}{I(H\beta)} \simeq O_{23}
\end{equation}

This calibration allows to derive metallicities using the theoretical ratios between
H$\beta$ and the other bright hydrogen recombination lines at bluer wavelengths, 
like H$\gamma$ and H$\delta$. It is therefore possible to derive metallicities of galaxies 
up to redshift $z\sim 1.25$ with the optical VIMOS spectra obtained by the VVDS. 
In that case, the O$_{2Ne3}$ parameter takes the following expression:
\begin{equation}
O_{2Ne3} = \frac{I([OII]) + 15.37 \cdot I([NeIII])}{I(Hi)}\cdot\frac{I(Hi)}{I(H\beta)}
\end{equation}

We keep the factor 15.37 obtained in P\'erez-Montero et al. (2007) for a sample
of HII regions of the local universe once the observed fluxes have been dereddened.
As shown in Figure \ref{o3ne3}, we obtain a very similar factor to correct 
the flux of the involved emission lines, with a reddening equivalent to an extinction
of A$_V$=0.5 mag. 
This correction has a very low impact on the determination of metallicities obtained with this
method: they are only 0.03 dex lower for those derived using H$\gamma$ and 0.01 dex lower using H$\delta$. 
Contrary to Paper I where equivalent widths were used as proposed by Kobulnicky \& Phillips (2003) and
Lamareille et al. (2006a), emission line fluxes work better in the case of O$_{2Ne3}$ parameter because the variations
of the continuum are more sensible to the underlying stellar population in this part of the spectrum.
To illustrate this point we have done an analysis of
the star-forming galaxies of the DR4 SDSS catalogue (Kauffman et al., 2003) whose involved emission lines have a signal-to-noise
larger than 5. We show in Figure \ref{ewa} the histograms of the residuals to the oxygen abundances derived from O$_{2Ne3}$ 
and whose emission lines have been corrected for reddening using the constants derived from the Balmer decrement. 
The filled histogram represents the residuals in relation with the abundances using the same parameter, but with
an average reddening correction for all the sample equivalent to A$_V$ = 0.5 mag. On the other hand, the empty thick line
histogram represents the residuals with the abundances deduced with the same parameter, but using equivalent widths and
correction factors based on the shape of the continuum calculated in a similar way as in Kobulnicky \& Phililps (2003). 
As we can see, both the mean value of the residuals and the dispersion of the distribution are much better in the first case.

As described in P\'erez-Montero et al. (2007), the O$_{2Ne3}$ parameter has the
same problems than O$_{23}$, the most important being its degeneracy with metallicity. 
This means that for a given value of the O$_{2Ne3}$ parameter we can find two possible values of the metallicity.
Among the numerous calibrations of the O$_{23}$ parameter there is always a calibration for high
metallicities (upper branch) and another one for low metallicities (lower branch) ({\em e.g.} Skillman, 1989).
It is thus mandatory to be able to distinguish between these two metallicity regimes. The ratio [Ne{\sc iii}]/[O{\sc ii}]
 has been proposed by Nagao et al. (2006) as an empirical calibrator of the metallicity but, due to its
high dependence on ionization parameter, it does not allow to retrieve accurate determinations of O/H.
However, it can be used to choose the appropriate branch (low or high metallicity regime) for the O$_{2Ne3}$ parameter.
The results obtained by P\'erez-Montero et al. (2007) indicate that galaxies with log([Ne{\sc iii}]/[O{\sc ii}]) $< -1.0$
are in the high metallicity regime, although the contrary is uncertain. In Figure \ref{logU}, we show the
relation between O$_{2Ne3}$ parameter and log([Ne{\sc iii}]/[O{\sc ii}]) for the VVDS galaxies harbouring 
the involved emission lines. We see that only a fraction of the objects lie 
in the high metallicity regime using this criterion.
For the rest of galaxies with log([Ne{\sc iii}]/[O{\sc ii}]) $> -1.0$, we constrain the metallicity regime by comparing the derived metallicities in low- and high-metallicity branches to the value derived with the Bayesian method described in Paper I or with the O$_{23}$ parameter in the case of the redshift bin $0.5 < z < 0.9$. After this comparison we conclude that 95\% of the sample of star-forming galaxies lie in the high metallicity regime.
The selected objects have values of O$_{2Ne3}$ in the range 0.0-1.5, corresponding to oxygen abundances in the 
range $8.0 <$ 12+log(O/H) $< 9.2$ in the upper branch.

In order to be consistent with metallicity estimates performed in Paper I, we use the calibrations
of the upper branch of O$_{23}$ parameter based on the models of Charlot \& Longhetti (2001).
As it has been shown by Kewley \& Ellison (2008), it is very important to keep the consistency between 
the different empirical calibrators used to derive the metallicity of galaxies. If not, systematic differences 
as high as 0.7 dex can be observed for the M-Z relation in the local universe as derived from the SDSS.

We check the consistency of this new method at higher redshift by comparing the metallicities 
derived using the O$_{2Ne3}$ and O$_{23}$ parameters in the VVDS star-forming galaxies at $z < 0.89$. 
Up to this redshift, all the emission lines required to compute these two parameters are observed 
in the VIMOS optical spectra.  For this purpose we take into account only the oxygen abundances derived 
with an accuracy better than 0.3 dex. The results of this comparison are shown in Figure \ref{ohcomp}. 
We distinguish three cases where O$_{2Ne3}$ parameter is estimated using H$\beta$ (268 galaxies), H$\gamma$ 
(217 galaxies) and H$\delta$ (187 galaxies) respectively. Dashed lines represent the 1:1 relation while the 
solid lines are the best linear fit to the data.
The best correlation is found in the first case, where the average of the residuals is 0.01 dex with a standard
deviation of 0.17 dex. For the second case (middle panel) the mean value of the residuals is also 0.01 dex, but the standard 
deviation increases to 0.25 dex. Finally, in the third case (right panel), we obtain a mean value of the residuals of 0.04 dex 
and a standard deviation of 0.24 dex. The same kind of analysis comparing the metallicities derived from the O$_{2Ne3}$ parameter 
on one side and the Bayesian Method on the other side gives a mean value of the residuals of 0.03 dex only with a standard deviation 
of 0.31 dex.

\begin{table*}

\caption{Sample of 42 star-forming VVDS galaxies in the redshift regime $0.89 < z < 1.24$ with a determination of both the stellar mass and the oxygen abundance. We report also their spectroscopic redshift and $I$-band (AB system) apparent magnitude.}
\label{list}
\begin{center}
\begin{tabular}{lcccc|lcccc}
\hline
\hline
Object & I$_{AB}$ & z & 12+log(O/H) & log(M$_*$/M$_\odot$) & Object & I$_{AB}$ & z & 12+log(O/H) & log(M$_*$/M$_\odot$) \\
\hline
VVDS000016597 & 23.24 & 0.9689 & 8.68 & 9.24$\pm$0.30 & VVDS020386743 & 22.58 & 1.0465 & 8.91 & 9.43$\pm$0.11 \\
VVDS000025365 & 22.80 & 0.8956 & 8.50 & 10.10$\pm$0.25 & VVDS020389342 & 23.76 & 0.9057 & 8.36 & 9.30$\pm$0.14 \\
VVDS000028748 & 21.24 & 0.9523 & 9.16 & 9.42$\pm$0.29 & VVDS020460957 & 23.36  & 0.9615 & 8.96 & 9.34$\pm$0.08 \\
VVDS020098524 & 23.13 & 0.9044 & 8.68 & 8.95$\pm$0.07 & VVDS020461893 & 23.45 & 1.0486 & 8.73 & 9.27$\pm$0.07 \\
VVDS020109569 & 22.50 & 1.0800 & 8.51 & 9.40$\pm$0.05 & VVDS020462475 & 22.76 & 0.9626 & 8.35 & 9.36$\pm$0.07 \\
VVDS020127928 & 21.23 & 0.9610 & 8.89 & 10.43$\pm$0.07 & VVDS220005349 & 22.36 & 0.9610 & 9.03 & 10.20$\pm$0.14 \\
VVDS020142000 & 22.16 & 0.9606 & 8.52 & 9.45$\pm$0.04 & VVDS220077555 & 22.10 & 0.9825 & 8.25 & 9.89$\pm$0.07 \\
VVDS020150832 & 23.19 & 0.9628 & 8.38 & 9.16$\pm$0.06 & VVDS220102663 & 21.83 & 1.1348 & 9.01 & 10.14$\pm$0.11 \\
VVDS020155109 & 22.17 & 0.9840 & 8.57 & 9.62$\pm$0.06 & VVDS220108994 & 22.3 & 0.9007 & 8.16 & 9.36$\pm$0.11 \\
VVDS020165728 & 23.11 & 0.9051 & 8.62 & 9.06$\pm$0.05 & VVDS220156608 & 22.06 & 0.9871 & 8.68 & 10.53$\pm$0.11 \\
VVDS020175617 & 23.18 & 0.9046 & 8.72 & 8.94$\pm$0.16 & VVDS220294849 & 22.17 & 0.9933 & 8.65 & 9.93$\pm$0.17 \\
VVDS020196604 & 23.12 & 0.9607 & 8.60 & 9.06$\pm$0.14 & VVDS220376206 & 21.85 & 1.2440 & 8.74 & 10.02$\pm$0.08 \\
VVDS020199214 & 22.62 & 0.9126 & 8.27 & 9.28$\pm$0.07 & VVDS220380153 & 21.55 & 0.9332 & 8.72 & 9.94$\pm$0.06 \\
VVDS020201361 & 23.27 & 1.0572 & 8.83 & 9.03$\pm$0.14 & VVDS220390996 & 21.76 & 0.9071 & 8.41 & 9.75$\pm$0.09  \\
VVDS020210745 & 23.97 & 0.8983 & 8.61 & 9.31$\pm$0.14 & VVDS220457904 & 22.30 & 1.0666 & 8.48 & 10.06$\pm$0.08 \\
VVDS020247631 & 23.37 & 0.9171 & 9.04 & 9.03$\pm$0.10 & VVDS220485833 & 20.99 & 0.9807 & 8.58 & 10.33$\pm$0.10 \\
VVDS020276040 & 21.71 & 0.9249 & 9.02 & 10.05$\pm$0.07 & VVDS220535047 & 22.29 & 1.0159 & 8.67 & 10.01$\pm$0.10 \\
VVDS020294045 & 22.80 & 1.0019 & 8.29 & 9.17$\pm$0.06 & VVDS220540780 & 21.91 & 0.9089 & 8.72 & 10.85$\pm$0.14 \\
VVDS020304000 & 23.46 & 0.9561 & 8.64 & 9.28$\pm$0.13 & VVDS220559147 & 22.25 & 0.8984 & 8.99 & 10.48$\pm$0.14 \\
VVDS020322281 & 22.89 & 0.9320 & 8.65 & 9.32$\pm$0.07 & VVDS220603467 & 21.85 & 0.9156 & 8.81 & 10.31$\pm$0.08 \\
VVDS020385905 & 22.36 & 0.9759 & 8.44 & 9.84$\pm$0.01 & VVDS220605155 & 22.06 & 0.9519 & 8.52 & 9.78$\pm$0.07 \\
\hline

\end{tabular}
\end{center}
\end{table*}

   \begin{figure*}
   \centering
   \includegraphics[width=15cm,clip=]{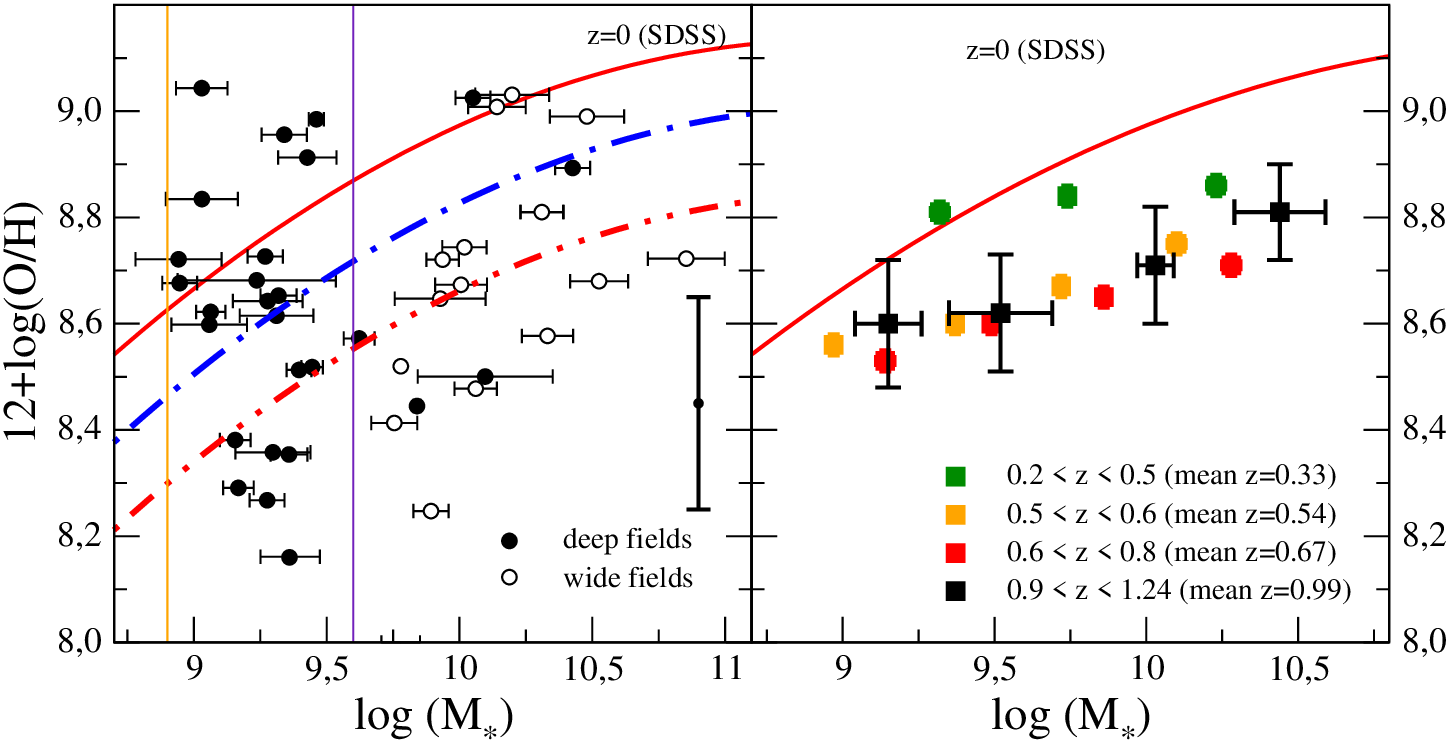}
   \caption{Relation between the gas-phase metallicity and the stellar mass (in solar masses) for VVDS star-forming galaxies in the redshift range 
$0.89 < z <  1.24$. In left panel, we show the 42 individual VVDS star-forming galaxies (black circles: deep fields; white circles: wide fields). 
Vertical solid lines represent the corresponding limiting mass to the deep and wide samples.
The solid red line represents the Tremonti et al. (2004) M-Z relation for the SDSS sample. 
We show as well the typical uncertainty of the O$_{2Ne3}$ parameter as vertical error bars.
The blue dashed-point curve represents the fit of the Tremonti curve to the VVDS galaxies of the deep sample, and
the red for the wide sample. In right panel, we show the running medians of the whole sample (black squares, both deep and wide fields), as compared 
with the results from Paper I at lower redshifts (coloured squares).}
              \label{MZ}
    \end{figure*}

\section{The mass-metallicity relation}

We study the M-Z relation for the redshift bin $0.89 < z < 1.24$, using the stellar masses and the metallicities 
derived in previous sections. We take only galaxies with a stellar mass higher than the
limiting mass imposed in Figure~\ref{zM} for a mean redshift $z=1$, corresponding to a logarithmic stellar mass in solar masses
of 8.9 for the deep fields and 9.6 for the wide fields.
We take only galaxies with an uncertainty in the calculation of the oxygen abundance lower than 0.3 dex, so
they remain 42 star-forming VVDS galaxies: 27 in the deep fields and 15 in the wide fields.
The galaxies are listed in Table \ref{list} along with their $I_{AB}$ apparent magnitude, spectroscopic redshift,
and derived metallicities and stellar masses. In the case of metallicity, all the errors due to
the uncertainties on the emission-line fluxes are lower than the intrinsic uncertainty of the empirical
parameter based on [Ne{\sc iii}] emission lines, which is 0.2 dex for the calibration of the upper branch
of the O$_{2Ne3}$ parameter (P\'erez-Montero et al., 2007). The spectra of all these galaxies are available
via the CENCOS database\footnote{\tt http://cencosw.oamp.fr}. We have to notice that, after applying the selection criteria 
described above, no galaxies from the 10h and 14h VVDS fields remain in the final sample.

In left panel of Figure~\ref{MZ} we show the relation between stellar mass and oxygen abundance for these
42 galaxies. For comparison, we  show also the M-Z relation derived in the local universe ($z < 0.3$) 
by Tremonti et al. (2004) with SDSS data (solid line). 
Note that the comparison can be made directly as the same method/calibration (namely Charlot \& Longhetti 2001, see sect. 2.3) has been used to derive metallicities in the local and high-redshift galaxies. 
The respective limiting mass for both deep and wide fields are shown as vertical solid lines. 

In order to test the global evolution in metallicity of star-forming galaxies we make the same analysis carried out 
in Paper I, consisting in the calculation of the mean shift on metallicity by fitting the same curve deduced for 
the SDSS sample to our data points and, hence, the derivation of a different zero-point.
These fits are shown in left panel of Figure~\ref{MZ} as a blue dashed-point line for the galaxies in the deep fields and
a red dashed-point line for the wide field galaxies. The mean difference in metallicity found with respect to the local M-Z relation 
are -0.29$\pm$0.21 and -0.12$\pm$0.15 for the wide and deep field galaxies respectively. As it was shown in Paper I, we see 
a stronger evolution of the mean metallicity of galaxies in the wide sample. As the limiting
mass of the wide sample is much higher than in the deep sample, this could be due to a a more 
significant evolution in metallicity for the most massive galaxies. The M-Z relation derived in the redshift range 
$0.89 < z < 1.24$ is rather similar to the one derived in Paper I for the redshift bin $0.7 < z < 0.9$.  
However, this difference between the wide and deep samples could be due to different selection effects. First the exposure time was higher for the deep fields which could lead to a higher dispersion in the metallicity distribution of these objects, as the number of high-metallicity galaxies, i.e. with fainter emission lines, is larger. There is also a larger number of low-mass galaxies in the deep sample, for which the metallicity is also lower compared with the average value.

We assess also the shape of the M-Z relation calculating a running median of the metallicity for the whole sample (both deep and 
wide fields) along the stellar mass axis in order to compare it with the results from Paper I and the relation deduced
for the local universe using the SDSS sample (Tremonti et al. 2004). 
The method to calculate these median is identical to that described in Paper I: we have taken four bins
in stellar mass, with a similar number of objects in each bin ($\approx$ 11) and then we calculate the median
of 1000 bootstraps estimates.
These medians are shown in the right panel of Figure~\ref{MZ} as black squares.
The error bars associated to the median metallicities of galaxies with $z>0.9$ are much higher than the ones derived 
in Paper I at lower redshifts, due to the lower number of objects in each mass bin.
The M-Z relation found for $0.9<z<1.24$ is fully compatible with the results of Paper I for $z>0.5$ in absolute value
and shape, pointing to a significant global chemical evolution for redshifts between 0 and 0.5.
However, for a redshift $z>0.5$, we do not find a significant global
evolution within the error bars up to $z = 1$.
Regarding the shape of the M-Z relation, we see a flattening as compared the with the M-Z relation in the local universe.
This could be a consequence of more effective yields for the more massive galaxies, 
where the evolution is stronger. However, again, we do not appreciate any significant evolution of the slope of the 
M-Z relation between redshifts 0.5 and 1.

It is expected somehow that the sample of star-forming galaxies considered in this work can be affected by some selection effects which lead to a higher fraction of objects with low metallicities. Firstly, the selection criterion on [Ne{\sc iii}] emission lines biases the sample towards lower metallicities as pointed by Nagao et al. (2006) and P\'erez-Montero et al. (2007). Secondly, the criterion on hydrogen recombination lines (H$\gamma$ and H$\delta$) implies a sample of galaxies with higher star-formation efficiencies. According to Ellison et al. (2008), this leads also to galaxies with lower metallicities. Nevertheless, the consistency between the metallicities derived using the O${_2Ne3}$ parameter in this redshift range and those at lower redshifts indicates that the probable selection effects inherent to our sample do not have a significant weight in the overall deduced metallicity.

\section{Summary and conclusions}

We have carried out a continuation of our study of the mass-metallicity relation of star-forming VVDS galaxies 
extending our analysis up to redshift $z \sim 1.24$.

We selected galaxies with a reliable measurement of [O{\sc ii}], [Ne{\sc iii}] and
H$\gamma$ or H$\delta$ emission lines and we identified star-forming galaxies with the diagnostic diagram based
on these lines and presented in P\'erez-Montero et al. (2007).  Then, we derived oxygen gas-phase
abundances using the empirical parameter O$_{2Ne3}$. We have shown that these new methods 
are consistent with those based on [O{\sc iii}] if we use the same calibrations.
As it is based on weaker emission lines that, generally, are more affected by absorption effects, 
the only problem appears as a selection effect, biaising the sample towards galaxies with bright 
emission lines and, hence, towards lower metallicities. 
The relative number of narrow-line AGNs at higher redshifts is thus to be taken with caution.

The final sample used to derive the relation between metallicity and stellar mass consists of 42 
VVDS star-forming galaxies with uncertainties lower than 0.3 dex in both oxygen abundance
and stellar mass. These have been selected to be 
larger than the previously calculated limiting mass for this redshift range.
Since the number of selected objects is sensibly lower than in our previous study (Paper I), the sample is not
as statistically significant but we have found fully consistent results.

The zero-point and the slope of the M-Z relation in the $0.9 < z < 1.24$ redshift range for both
the deep and wide fields are rather similar 
to the values reported in Paper I for $z>0.5$, showing a significant global cosmic chemical evolution
in relation with the local universe. We detect also a flattening of the M-Z relation in comparison with the 
SDSS-based relation in the local universe (Tremonti et al., 2004).
This evolution is stronger for the galaxies of the wide sample, consistently with the 
fact that most massive galaxies have undergone a stronger chemical evolution.
The evolution of massive galaxies compared with the local sample can be understood as a
consequence of lower effective yields for the most massive galaxies at larger redshifts, according
to an hierarchical scenario of formation and evolution of galaxies.

Nevertheless, despite the probable selection effects of the final sample,
we do not appreciate any significant evolution neither in the absolute value of the metallicity
nor in the shape of the M-Z relation for redshifts larger than 0.5. This can be a consequence of the smaller
time interval in the redshift range $0.7<z<1$ combined with a larger uncertainty due to the lower number of
objects with an appropriate signal-to-noise ratio for the used emission-lines. In any case, our results
are fully compatible with those published for $z \approx 1$ in the DEEP2 sample (Shapley et al.,
2005; Liu et al., 2008) or in the GOODS sample (Cowie \& Barger, 2008), 
who find 0.2-0.3 dex lower metallicities in relation with the local universe.

\begin{acknowledgements}
We thank C. Tremonti for having made the original {\em platefit} code available to the VVDS collaboration.
 We also thank the referee, Roberto Maiolino, for many valuable suggestions and comments, which have helped us to improve this paper.\\
This work has been partially supported by the CNRS-INSU and its Programmes Nationaux de Galaxies et de Cosmologie (France)
and by project AYA-2004-08260-C03-03 of the Spanish National Plan for Astronomy and Astrophysics.\\
The VLT-VIMOS observations have been carried out on guarateed time (GTO) allocated by the European
Southern Observatory (ESO) to the VIRMOS consortium, under a contractual agreement between the Centre
National de la Recherche Scientifique of France, heading a consortium of French and Italian institutes,
and ESO, to design, manufacture and test the VIMOS instrument.\\
Based on observations obtained with MegaPrime/MegaCam, a joint project of CFHT and CEA/DAPNIA, at the Canada-France-
Hawaii Telescope (CFHT) which is operated by the National Research Council (NRC) of Canada, the Institut
National des Science de l'Univers of the Centre National de la Recherche Scientifique (CNRS) of France and
the University of Hawaii. This work is based in part on data products produced at TERAPIX and the Canadian
Astronomy Data Centre as part of the Canada-France-Hawaii Telescope Legacy Survey, a collaboration
project of NRC and CNRS.
\end{acknowledgements}

\noindent $^1$ Laboratoire d'Astrophysique de Toulouse-Tarbes, Universit\'e de Toulouse, CNRS, Observatoire Midi-Pyr\'en\'ees, 14 av. E. Belin, F-31400, Toulouse, France \\
$^2$ IASF-INAF, Via Bassini 15, I-20133, Milano, Italy \\
$^3$ INAF-Osservatorio Astronomico di Bologna, Via Ranzani 1, I-40127, Bologna, Italy \\
$^4$ IRA-INAF, Via Gobetti 101, I-40129, Bologna, Italy \\
$^5$ INAF-Osservatorio Astronomico di Capodimonte, Via Moiariello 16, I-80131, Napoli, Italy \\
$^6$ Universit\`a di Bologna, Dipartimento di Astronomia, Via Ranzani 1, I-40127, Bologna, Italy \\
$^7$  Laboratoire d'Astrophysique de Marseille, UMR 6110
CNRS-Universit\'e de Provence, 38 rue Frederic Joliot-Curie, F-13388
Marseille Cedex 13, France \\
$^8$ Max Planck Institut f\"ur Astrophysik, D-85741, Garching, Germany \\
$^9$ INAF-Osservatorio Astronomico di Brera, Via Brera 28, I-20021, Milan, Italy \\
$^{10}$ Institut d'Astrophysique de Paris, UMR 7095, 98 bis Bvd Arago, F-75014, Paris, France \\
$^{11}$ Observatoire de Paris, LERMA, 61 Avenue de l'Observatoire, F-75014, Paris, France \\
$^{12}$ Astrophysical Institute Potsdam, An der Sternwarte 16, D-14482, Potsdam, Germany \\
$^{13}$ INAF-Osservatorio Astronomico di Roma, Via di Frascati 33, I-00040, Monte Porzio Catone, Italy \\
$^{14}$ Universit\'a di Milano-Bicocca, Dipartimento di Fisica, Piazza delle Scienze 3, I-20126, Milano, Italy \\
$^{15}$ Integral Science Data Centre, ch. d'\'Ecogia 16, CH-1290, Versoix, Switzerland \\
$^{16}$ Geneva Observatory, ch. des Maillettes 51, CH-1290, Sauverny, Switzerland \\
$^{17}$ Astronomical Observatory of the Jagiellonian University, ul Orla 171, PL-30-244, Krak{\'o}w, Poland \\
$^{18}$ Centre de Physique Th\'eorique, UMR 6207 CNRS-Universit\'e de Provence, F-13288, Marseille, France \\
$^{19}$ Centro de Astrof{\'{i}}sica da Universidade do Porto, Rua das Estrelas, P-4150-762, Porto, Portugal \\
$^{20}$ Institute for Astronomy, 2680 Woodlawn Dr., University of Hawaii, Honolulu, Hawaii, 96822, USA \\
$^{21}$ School of Physics \& Astronomy, University of Nottingham, University Park, Nottingham, NG72RD, UK \\
$^{22}$ Max Planck Institut f\"ur Extraterrestrische Physik (MPE), Giessenbachstrasse 1,
D-85748 Garching bei M\"unchen,Germany \\
$^{23}$ Canada France Hawaii Telescope corporation, Mamalahoa Hwy,  
Kamuela, HI-96743, USA

\end{document}